\documentclass{PoS}

\usepackage{ifthen} 
\newboolean{pdflatex}
\setboolean{pdflatex}{true} 
\newboolean{articletitles}
\setboolean{articletitles}{true} 
\newboolean{uprightparticles}
\setboolean{uprightparticles}{false} 
\newboolean{inbibliography}
\setboolean{inbibliography}{false} 

\usepackage{xspace} 
\usepackage{upgreek}

\def\MagUp {\mbox{\em Mag\kern -0.05em Up}\xspace}

\ifthenelse{\boolean{uprightparticles}}%
{

 \def\Ppi         {\ensuremath{\uppi}\xspace}

 \def\PDelta      {\ensuremath{\Delta}\xspace}                 
 \def\PXi      {\ensuremath{\Xi}\xspace}                 
 \def\PLambda      {\ensuremath{\Lambda}\xspace}                 
 \def\PSigma      {\ensuremath{\Sigma}\xspace}                 
 \def\POmega      {\ensuremath{\Omega}\xspace}                 
 \def\PUpsilon      {\ensuremath{\Upsilon}\xspace}                 
 
 \def\PB      {\ensuremath{\mathrm{B}}\xspace}                 
                  
 \def\PD      {\ensuremath{\mathrm{D}}\xspace}

 \def\PK      {\ensuremath{\mathrm{K}}\xspace}

 \def\Pb      {\ensuremath{\mathrm{b}}\xspace}

 \def\Pi      {\ensuremath{\mathrm{i}}\xspace}

 \def\Pp      {\ensuremath{\mathrm{p}}\xspace}

 \def\Ps      {\ensuremath{\mathrm{s}}\xspace}

}
{

 \def\Ppi         {\ensuremath{\pi}\xspace}

 \mathchardef\PDelta="7101
 \mathchardef\PXi="7104
 \mathchardef\PLambda="7103
 \mathchardef\PSigma="7106
 \mathchardef\POmega="710A
 \mathchardef\PUpsilon="7107
                  
 \def\PB      {\ensuremath{B}\xspace}                 
                  
 \def\PD      {\ensuremath{D}\xspace}

 \def\PK      {\ensuremath{K}\xspace}

 \def\Pb      {\ensuremath{b}\xspace}

 \def\Pi      {\ensuremath{i}\xspace}

 \def\Pp      {\ensuremath{p}\xspace}

 \def\Ps      {\ensuremath{s}\xspace}

}

\makeatletter
\ifcase \@ptsize \relax
  \newcommand{\miniscule}{\@setfontsize\miniscule{4}{5}}
\or
  \newcommand{\miniscule}{\@setfontsize\miniscule{5}{6}}
\or
  \newcommand{\miniscule}{\@setfontsize\miniscule{5}{6}}
\fi
\makeatother

\DeclareRobustCommand{\optbar}[1]{\shortstack{{\miniscule (\rule[.5ex]{1.25em}{.18mm})}
  \\ [-.7ex] $#1$}}

\def\squark    {{\ensuremath{\Ps}}\xspace}

\def\bquark    {{\ensuremath{\Pb}}\xspace}

\def\pion   {{\ensuremath{\Ppi}}\xspace}
\def\piz    {{\ensuremath{\pion^0}}\xspace}

\def\pip    {{\ensuremath{\pion^+}}\xspace}
\def\pim    {{\ensuremath{\pion^-}}\xspace}
\def\pipm   {{\ensuremath{\pion^\pm}}\xspace}

\def\kaon    {{\ensuremath{\PK}}\xspace}
  \def\Kbar    {{\kern 0.2em\overline{\kern -0.2em \PK}{}}\xspace}

\def\KorKbar    {\kern 0.18em\optbar{\kern -0.18em K}{}\xspace}

\def\Kp      {{\ensuremath{\kaon^+}}\xspace}
\def\Km      {{\ensuremath{\kaon^-}}\xspace}

\def\Kmp     {{\ensuremath{\kaon^\mp}}\xspace}
\def\KS      {{\ensuremath{\kaon^0_{\mathrm{ \scriptscriptstyle S}}}}\xspace}

\def\Kstar   {{\ensuremath{\kaon^*}}\xspace}
\def\KorKbarstar {\kern 0.18em\optbar{\kern -0.18em K}{}^*\xspace}

  \def\Dbar    {{\kern 0.2em\overline{\kern -0.2em \PD}{}}\xspace}
\def\D       {{\ensuremath{\PD}}\xspace}

\def\DorDbar    {\kern 0.18em\optbar{\kern -0.18em D}{}\xspace}
\def\Dz      {{\ensuremath{\D^0}}\xspace}
\def\Dzb     {{\ensuremath{\Dbar{}^0}}\xspace}
\def\Dp      {{\ensuremath{\D^+}}\xspace}
\def\Dm      {{\ensuremath{\D^-}}\xspace}

\def\Dstarzb {{\ensuremath{\Dbar{}^{*0}}}\xspace}
\def\Dstarp  {{\ensuremath{\D^{*+}}}\xspace}

\def\B       {{\ensuremath{\PB}}\xspace}
\def\Bbar    {{\ensuremath{\kern 0.18em\overline{\kern -0.18em \PB}{}}}\xspace}

\def\BorBbar    {\kern 0.18em\optbar{\kern -0.18em B}{}\xspace}
\def\Bz      {{\ensuremath{\B^0}}\xspace}
\def\Bzb     {{\ensuremath{\Bbar{}^0}}\xspace}
\def\Bu      {{\ensuremath{\B^+}}\xspace}

\def\Bp      {{\ensuremath{\Bu}}\xspace}

\def\Bs      {{\ensuremath{\B^0_\squark}}\xspace}

  \def\Y#1S{\ensuremath{\PUpsilon{(#1S)}}\xspace}

\def\proton      {{\ensuremath{\Pp}}\xspace}

\def\Lz          {{\ensuremath{\PLambda}}\xspace}
\def\Lbar        {{\ensuremath{\kern 0.1em\overline{\kern -0.1em\PLambda}}}\xspace}
\def\LorLbar    {\kern 0.18em\optbar{\kern -0.18em \PLambda}{}\xspace}

\def\Lb      {{\ensuremath{\Lz^0_\bquark}}\xspace}

\def\to                 {\ensuremath{\rightarrow}\xspace}

\def\CP                {{\ensuremath{C\!P}}\xspace}

\def\AT#1     {\ensuremath{A_{\mathrm{T}}^{#1}}\xspace}           

\def\C#1      {\ensuremath{\mathcal{C}_{#1}}\xspace}                       
\def\Cp#1     {\ensuremath{\mathcal{C}_{#1}^{'}}\xspace}                    
\def\Ceff#1   {\ensuremath{\mathcal{C}_{#1}^{\mathrm{(eff)}}}\xspace}        
\def\Cpeff#1  {\ensuremath{\mathcal{C}_{#1}^{'\mathrm{(eff)}}}\xspace}       
\def\Ope#1    {\ensuremath{\mathcal{O}_{#1}}\xspace}                       
\def\Opep#1   {\ensuremath{\mathcal{O}_{#1}^{'}}\xspace}                    

\newcommand{\tev}{\ifthenelse{\boolean{inbibliography}}{\ensuremath{~T\kern -0.05em eV}}{\ensuremath{\mathrm{\,Te\kern -0.1em V}}}\xspace}
\newcommand{\gev}{\ensuremath{\mathrm{\,Ge\kern -0.1em V}}\xspace}
\newcommand{\mev}{\ensuremath{\mathrm{\,Me\kern -0.1em V}}\xspace}
\newcommand{\kev}{\ensuremath{\mathrm{\,ke\kern -0.1em V}}\xspace}
\newcommand{\ev}{\ensuremath{\mathrm{\,e\kern -0.1em V}}\xspace}
\newcommand{\gevc}{\ensuremath{{\mathrm{\,Ge\kern -0.1em V\!/}c}}\xspace}
\newcommand{\mevc}{\ensuremath{{\mathrm{\,Me\kern -0.1em V\!/}c}}\xspace}
\newcommand{\gevcc}{\ensuremath{{\mathrm{\,Ge\kern -0.1em V\!/}c^2}}\xspace}
\newcommand{\gevgevcccc}{\ensuremath{{\mathrm{\,Ge\kern -0.1em V^2\!/}c^4}}\xspace}
\newcommand{\mevcc}{\ensuremath{{\mathrm{\,Me\kern -0.1em V\!/}c^2}}\xspace}

\def\invfb   {\ensuremath{\mbox{\,fb}^{-1}}\xspace}

\def\gsim{{~\raise.15em\hbox{$>$}\kern-.85em
          \lower.35em\hbox{$\sim$}~}\xspace}
\def\lsim{{~\raise.15em\hbox{$<$}\kern-.85em
          \lower.35em\hbox{$\sim$}~}\xspace}

\def\tell1  {TELL1\xspace}
\def\ukl1   {UKL1\xspace}

\usepackage{cite} 
\usepackage{mciteplus}
\usepackage{lineno}

\title{Current challenges and future prospects for $\gamma$ from $B\to D hh^\prime$ decays}

\ShortTitle{Current challenges and future prospects for $\gamma$ from $B\to D hh^\prime$ decays}

\author{
  \speaker{Tim Gershon} \thanks{on behalf of the LHCb collaboration}
  \\
  University of Warwick\\
  E-mail: \email{t.j.gershon@warwick.ac.uk}
} 

\abstract{
  Decays of the type $B\to D hh^\prime$, where a $b$~hadron decays to a neutral charm meson that can be an admixture of $\Dz$ and $\Dzb$ states together with two light particles that are typically a kaon and a pion, have demonstrated potential to enable precise determinations of the angle $\gamma$ of the CKM Unitarity Triangle.
  The current status and future prospects of these measurements are reviewed.
}

\FullConference{ 9th International Workshop on the CKM Unitarity Triangle\\
         28  November -- 3 December 2016\\
         Tata Institute for Fundamental Research (TIFR), Mumbai, India}

\begin{document}


The angle $\gamma$ of the CKM Unitarity Triangle is a key parameter of quark flavour physics.
It is the only \CP-violating parameter that can be measured using only tree-level decays, and as such is a benchmark Standard Model reference point (for a detailed review, see Ref.~\cite{Gershon:2016fda}).
Its precise determination is essential in order to be able to disentangle possible contributions from physics beyond the Standard Model to other \CP-violating observables that enter the global CKM fit. 

The channels that are most commonly used to determine $\gamma$ are of the type $B \to DK$, where a $b$~hadron decays to a neutral charm meson together with a kaon.
When the final state is accessible to both $\Dz$ and $\Dzb$ decays, the neutral \D meson is an admixture of the flavour eigenstates.
Since these are produced through $b \to u$ and $b \to c$ transitions, their interference is sensitive to the relative weak phase $\gamma$~\cite{Gronau:1990ra,Gronau:1991dp,Atwood:1996ci,Atwood:2000ck}.
By measuring the rates and \CP asymmetries of such decays, $\gamma$ can be determined with negligible theoretical uncertainty~\cite{Brod:2013sga}.

The method can be extended to $B \to DK\pi$ decays.
In this case, amplitude analysis of the $B$ decay Dalitz plot~\cite{Dalitz:1953cp} provides direct information about the relative phases, and therefore can be used to obtain precise information about $\gamma$ without ambiguities in the solution.
In particular, in the Dalitz plot analysis of $\Bz \to D\Kp\pim$ decays,\footnote{The inclusion of charge conjugate processes is implied throughout unless explicitly stated otherwise.} interference between $\Bz \to D\Kstar(892)^0$ and $\Bz \to D_2^*(2460)^- \Kp$ amplitudes can be used to obtain more information about $\gamma$ than is available in a quasi-two-body analysis~\cite{Gershon:2008pe,Gershon:2009qc}.
A key point is that the $D_2^*(2460)^- \Kp$ amplitude is flavour-tagged and therefore does not depend on the \D decay final state.
The method also allows the determination of additional hadronic parameters such as coherence factors that enter the formalism of the quasi-two-body approach~\cite{Gronau:2002mu}.

Determination of $\gamma$ with this method has recently been achieved, for the first time, by LHCb. 
In the first step, the Dalitz plot distribution of $\Bz \to \Dzb\Kp\pim$ decays is obtained by fitting a sample reconstructed in the $\Dzb \to \Kp\pim$ channel (which is flavour-specific, to a good approximation).
With the full LHC run~I data sample of $3 \invfb$ of $pp$ collision data at centre-of-mass energies of $\sqrt{s} = 7$ and $8 \tev$, $2344 \pm 66$ signal decays are found inside the \Bz\ signal window~\cite{LHCb-PAPER-2015-017}.
The Dalitz plot analysis provides a model for the $b \to c$ transition, and reveals that the largest resonant contributions are from the $\Kstar(892)^0$ and $D_2^*(2460)^-$ states, with additional significant components from $K\pi$ and $D\pi$ S-waves.
Results on the masses and widths of the $D_0^*(2400)^-$ and $D_2^*(2460)^-$ states are also obtained in the analysis.  

\begin{figure}[!tb]
  \centering
  \includegraphics[width=0.48\textwidth]{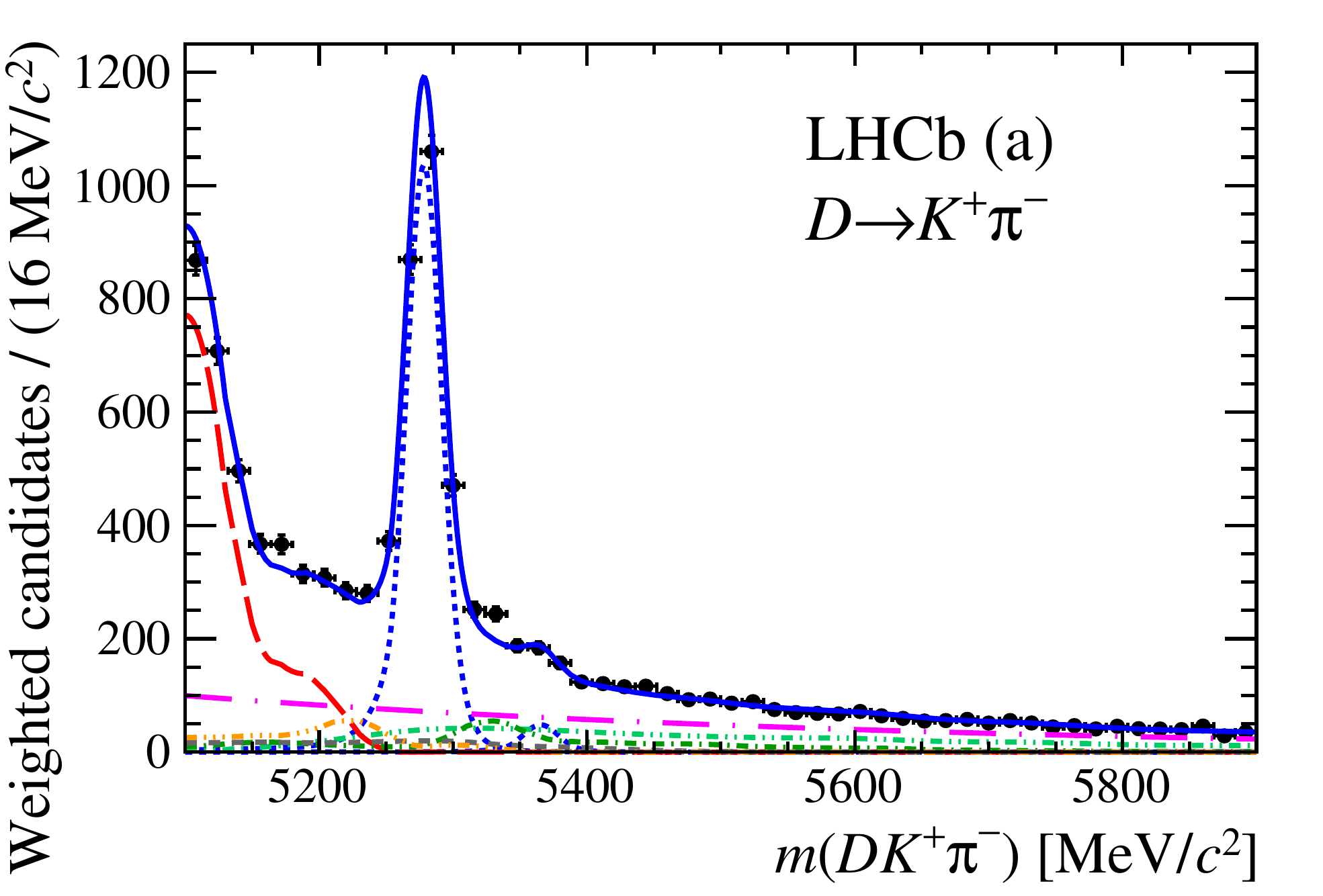}
  \includegraphics[width=0.48\textwidth]{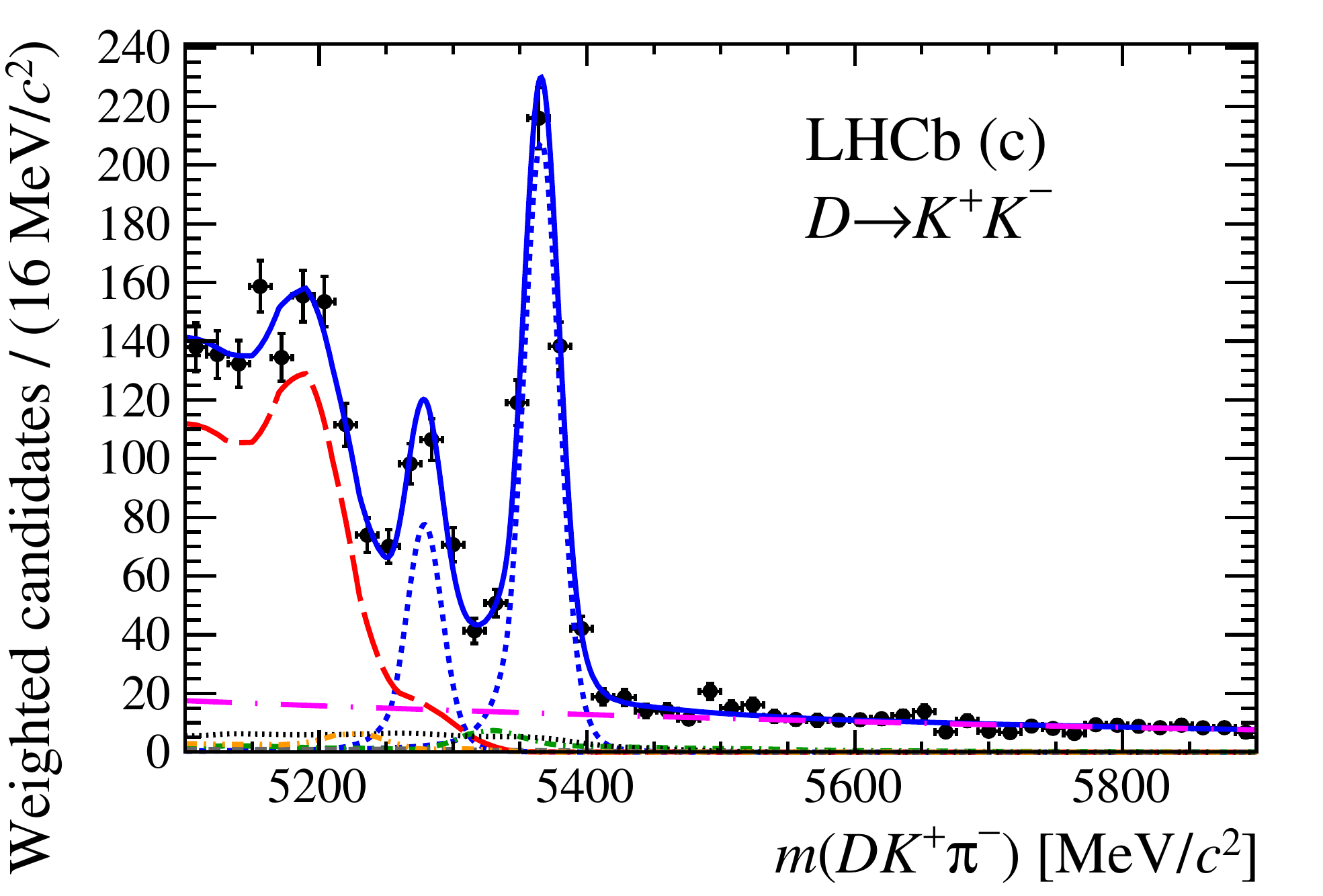} \\
  \includegraphics[width=0.48\textwidth]{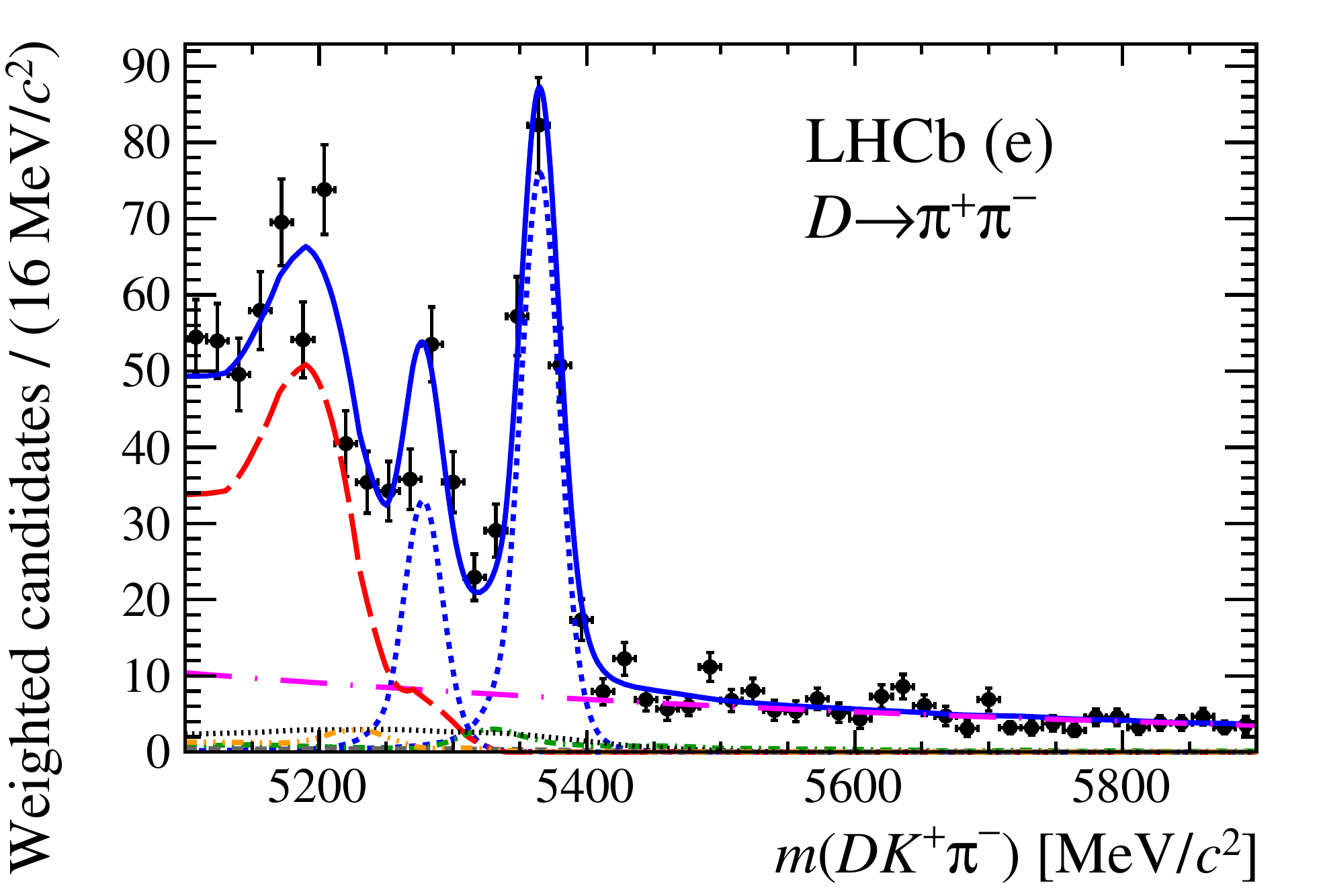}
  \raisebox{2cm}{\includegraphics[width=0.48\textwidth]{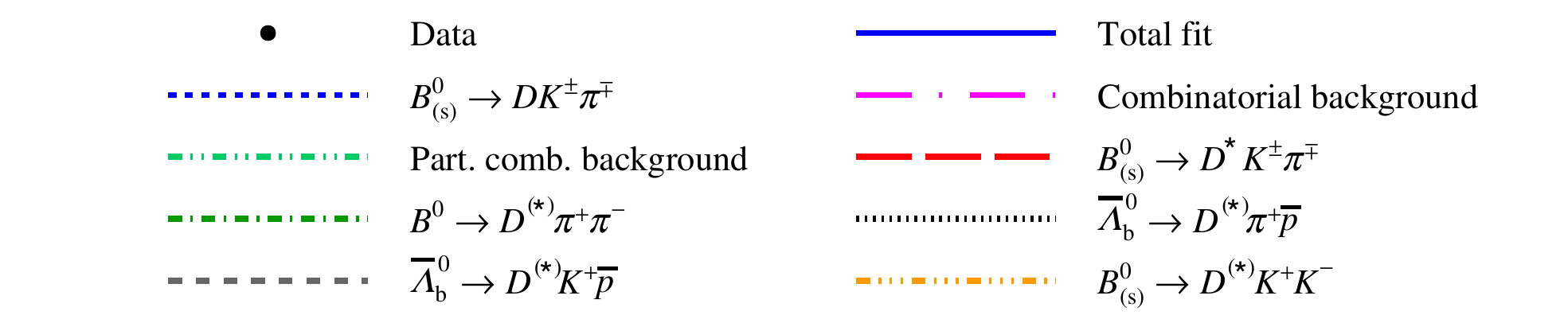}}
  \caption{
    Candidates for $\Bz \to D\Kp\pim$ decays in the (top left) $D\to\Kp\pim$, (top right) $\Kp\Km$ and (bottom left) $\pip\pim$ channels~\cite{LHCb-PAPER-2015-059}.
    The largest peak in the \CP-eigenstate modes is due to $\Bs \to \Dzb\Km\pip$ decays, with an associated satellite peak (long-dashed red line) from $\Bs \to \Dstarzb\Km\pip$ decays.
    The candidates have been weighted by the signal-to-background fractions in the different samples that are fitted.
  }
  \label{fig:DKpi-BmassFits}
\end{figure}

With the $b \to c$ model thus established, the analysis is extended to include decays of the \D meson to the \CP-even $\Kp\Km$ and $\pip\pim$ final states, where yields of $339 \pm 22$ and $168 \pm 19$ signal events are available inside the \Bz\ signal window~\cite{LHCb-PAPER-2015-059}, as shown in Fig.~\ref{fig:DKpi-BmassFits}.
A simultaneous Dalitz plot fit, implemented in Laura++~\cite{Laura++} with the {\it jFit} method~\cite{Ben-Haim:2014afa}, is carried out to the samples with $D\to\Kp\pim$, $\Kp\Km$ and $\pip\pim$ -- this is the first such simultaneous Dalitz plot analysis ever performed.
The $\Bz \to D\Kp\pim$, $D\to\Kp\pim$ sample is fitted with the $b \to c$ model, while the model is modified for the $D \to \Kp\Km$ and $\pip\pim$ samples to account for effect of the $b \to u$ contributions.
Specifically, the complex coefficient $c_j$ which describes the relative contribution of the resonance $j$ to the overall amplitude is unchanged for $D\pim$ resonances, since the charge of the pion tags the flavour of the resonance, while amplitudes for $\Kp\pim$ resonances receive additional contributions,
\begin{equation}
  \label{eq:param1}
  c_j \longrightarrow 
  \left\{ 
    \begin{array}{cl}
      c_j & {\rm for \ a} \ \D\pim \ {\rm resonance} \, ,\\
      c_j \left[ 1 + x_{\pm,\,j} + i y_{\pm,\,j} \right] & {\rm for \ a} \ \Kp\pim \ {\rm resonance} \, ,
    \end{array}
    \right.  
\end{equation}
with $x_{\pm,\,j} = r_{B,\,j} \cos\left(\delta_{B,\,j} \pm \gamma\right)$ and $y_{\pm,\,j} = r_{B,\,j} \sin\left(\delta_{B,\,j} \pm \gamma\right)$, where the $+$ and $-$ signs correspond to $\Bz$ and $\Bzb$ decay amplitudes, respectively.
Here $r_{B,\,j}$ and $\delta_{B,\,j}$ are the relative magnitude and strong phase of the $b \to u$ and $b \to c$ amplitudes for each $\Kp\pim$ resonance $j$.
A component corresponding to the $\Bz \to D_{s1}^*(2700)^+\pim$ decay, which is mediated by the $b \to u$ amplitude alone, is also included.

\begin{figure}[!tb]
  \centering
  \includegraphics[width=0.48\textwidth]{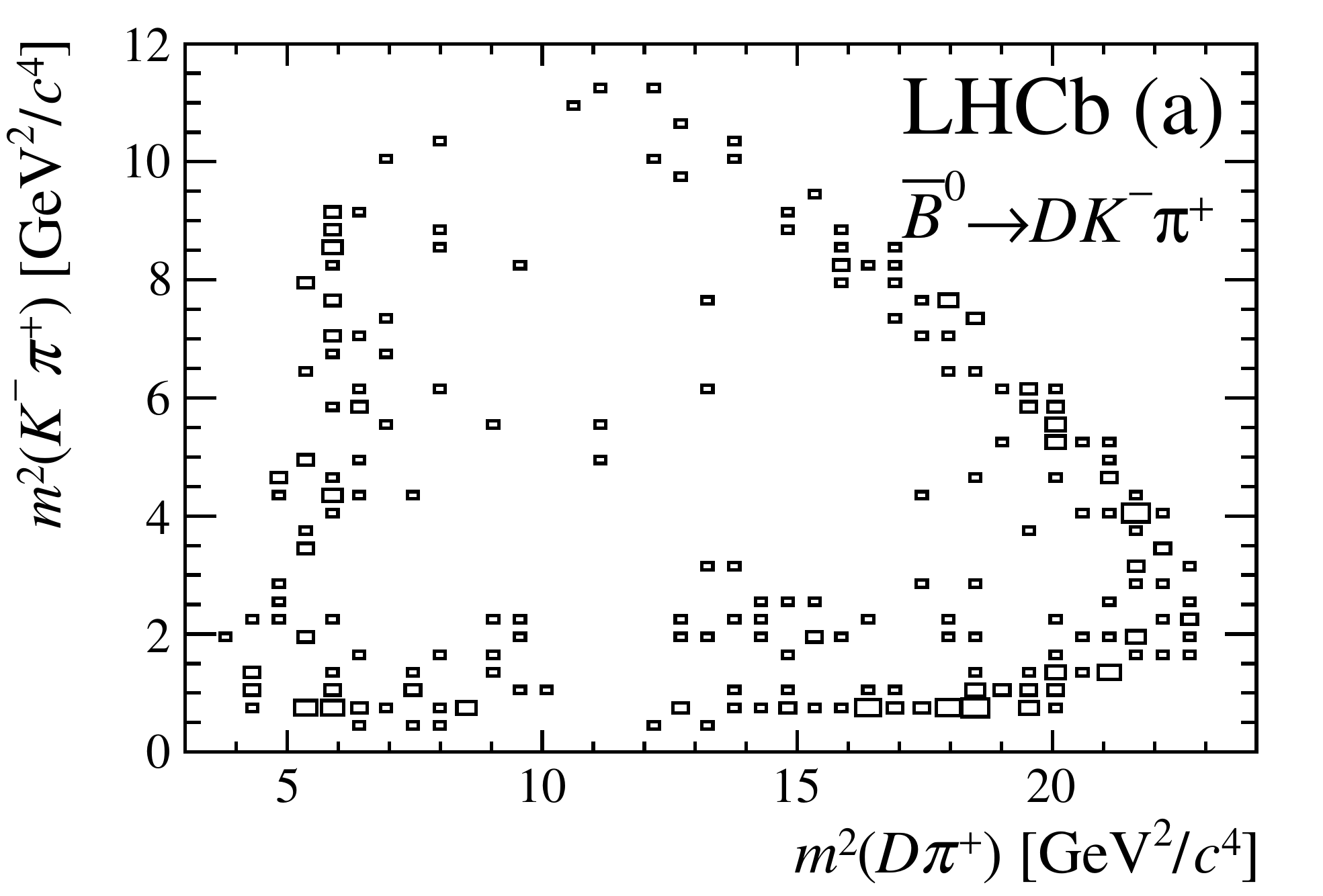}
  \includegraphics[width=0.48\textwidth]{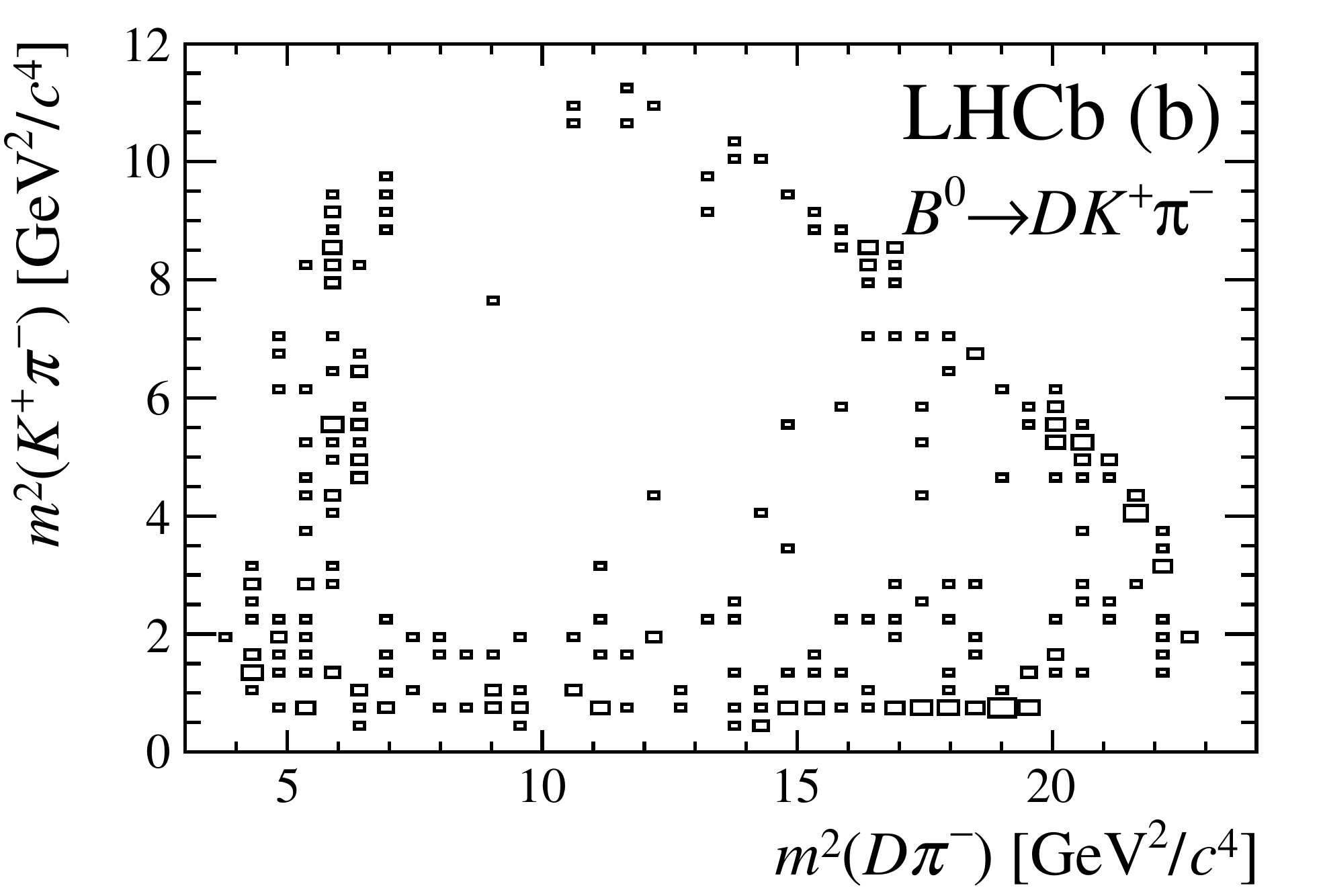} \\
  \includegraphics[width=0.48\textwidth]{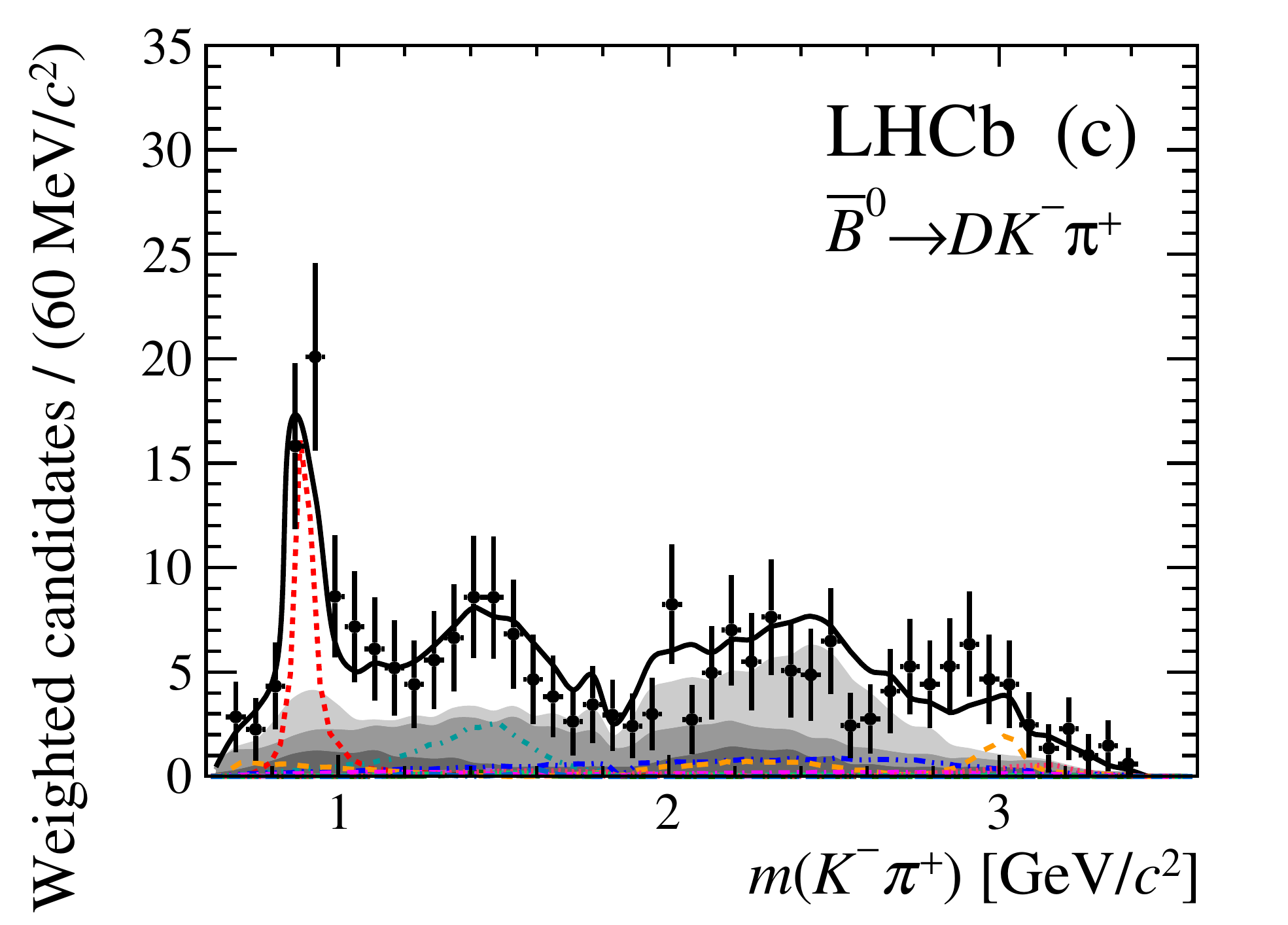}
  \includegraphics[width=0.48\textwidth]{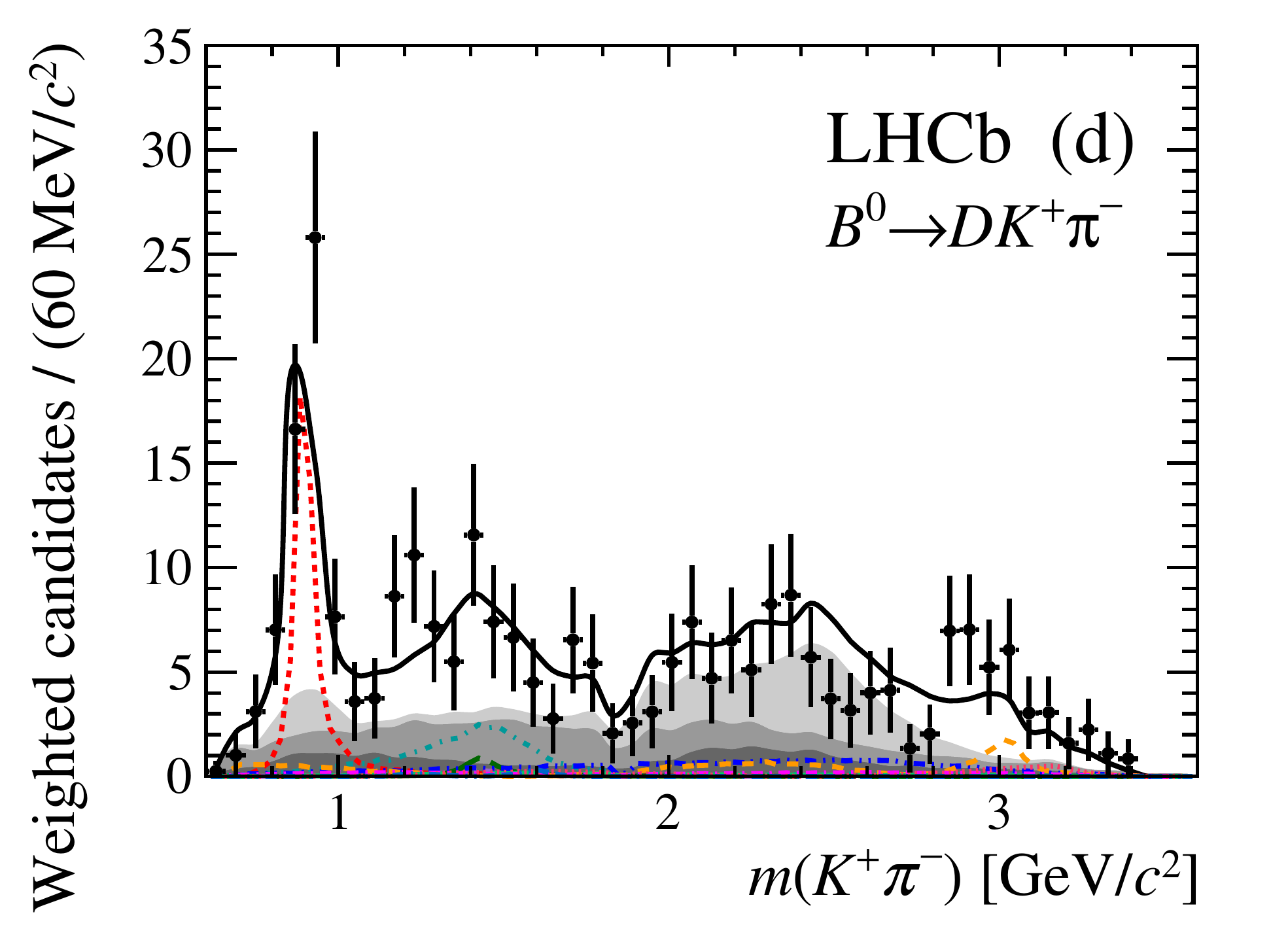}
  \caption{
    (Top) Dalitz plots for (left) $\Bzb$ and (right) $\Bz$ candidates, together with (bottom) their projections onto $m(\Kmp\pipm)$ with results of the fit superimposed~\cite{LHCb-PAPER-2015-059}.
    In the projections the shaded areas indicate backgrounds, while the red dotted line is the contribution from the $\KorKbarstar(892)^0$ resonance.
  }
  \label{fig:DKpi-DPFits}
\end{figure}

The Dalitz plots for candidates in the $\Bz \to D\Kp\pim$, $D \to \Kp\Km$ and $\pip\pim$ samples combined are shown in Fig.~\ref{fig:DKpi-DPFits}, together with projections of the data and the fit result onto $m(\Kmp\pipm)$.  
Within the available statistics, there is no evidence for \CP violation.
The results for the parameters of the $\Bz \to D\Kstar(892)^0$ decay are consistent with those of a quasi-two-body analysis based on the same data sample~\cite{LHCb-PAPER-2014-028}.
The determination of the $x_\pm, y_\pm$ parameters of Eq.~(\ref{eq:param1}) allows also a comparison with results obtained from the $\Bz \to D\Kstar(892)^0$, $D \to \KS \pip\pim$ and $\KS\Kp\Km$ mode~\cite{LHCb-PAPER-2016-006,LHCb-PAPER-2016-007}: the $x_\pm$ results from the $\Bz \to D\Kp\pim$ Dalitz plot analysis are slightly more precise, while the $y_\pm$ results are slightly less precise; all results are consistent.

Since the central values of the $x_\pm, y_\pm$ parameters are not significantly different from zero, limited precision on $\gamma$ is obtained using the results of the $\Bz \to D\Kp\pim$ Dalitz plot analysis alone.
However, the analysis also yields information about the hadronic parameters needed to interpret results obtained from quasi-two-body analyses.
In particular, the coherence factor $\kappa$, which would be unity in the case that the $\Kstar(892)^0$ selection window contains only contributions from the $\Kstar(892)^0$ resonance, is determined to be $\kappa = 0.958 \,^{+0.005}_{-0.010} \,^{+0.002}_{-0.045}$, where the uncertainties are statistical and systematic.
The results therefore have an important impact on the combined determination of $\gamma$ using results from all $B \to DK$ type processes~\cite{LHCb-PAPER-2016-032,Amhis:2016xyh}.
The LHCb combination~\cite{LHCb-PAPER-2016-032} gives for the ratio of magnitudes of $b \to u$ and $b \to c$ amplitudes, $r_B(D\Kstar(892)^0) = 0.218\,^{+0.045}_{-0.047}$, smaller than but consistent with the expected value of $\sim 0.3$. 
Analyses with larger data samples will therefore be important to see if this value increases, in which case $\Bz \to D\Kstar(892)^0$ decays will have an even larger impact on the overall combination than now.
In addition to increasing the size of the sample, it will be important to improve understanding on $K\pi$ and $D\pi$ S-wave amplitudes (for which good progress has been reported recently~\cite{LHCb-PAPER-2016-026,Palano:2017nex}) and to control background contributions from $\Bs \to \Dstarzb\Km\pip$ decays (the $\Bs \to \Dzb\Km\pip$ Dalitz plot has already been studied~\cite{LHCb-PAPER-2014-035,LHCb-PAPER-2014-036}).
More \D meson decay modes can also be added, including the possibility of a model-independent $\Bz \to D\Kp\pim$, $D\to \KS\pip\pim$ double Dalitz plot analysis~\cite{Gershon:2009qr}.

Given the success of the $B \to DK\pi$ Dalitz plot analysis, it is reasonable to ask whether similar approaches can be applied for other $B \to Dhh^\prime$ modes.
The isospin partner $\Bp \to D\Kp\piz$ would be more challenging experimentally, due to the presence of a neutral pion in the final state.
A further challenge in this channel is that $D\piz$ resonances are not flavour tagged by the charge of the pion, so the associated amplitudes can differ depending on the $D$ meson final state.  
While this complicates the formalism, it also means that in principle there may be more interference between $b \to u$ and $b \to c$ amplitudes, and therefore better sensitivity to $\gamma$.
Although this mode has been investigated in the past~\cite{Aleksan:2002mh}, a more detailed investigation taking into account the latest knowledge is warranted~\cite{Gershon-ICHEP}.

A further potential advantage of the $\Bp \to D\Kp\piz$ mode is that for many $D\pi$ resonances, the relative magnitude of the contributing amplitudes $r_B$ can be known independently from studies of the $\Bp \to \Dp\Kp\pim$ and $\Bp \to \Dm\Kp\pip$ decays~\cite{Sinha:2004ct}.
(This is not the case for the $D^*(2007)^0$ resonance, which is below threshold for decay to $\Dp\pim$, but has other advantages~\cite{Bondar:2004bi}.  The quasi-two-body approach is preferable for analysis of decays involving this narrow resonance.)
Both these modes have recently been observed by LHCb~\cite{LHCb-PAPER-2015-007,LHCb-PAPER-2015-054}.
A large $D_2^*(2460)^0$ component in seen in the Dalitz plot analysis of the favoured mode.
In the suppressed mode, the available statistics are not sufficient for amplitude analysis, so instead a novel method involving weighting data by angular moments is used to set a limit on the $\Dbar_2^*(2460)^0$ contribution.
These results give an upper limit $r_B(D_2^*(2460)\Kp) < 0.30 \ (0.36) \ {\rm at}\ 90 \ (95) \ \%$ confidence level.

Extending to four-body decays, similar methods could potentially be used to determine $\gamma$ from the interference of $b \to u$ and $b \to c$ amplitudes in $\Bp \to D_1(2420)\Kp$ decays.  
A possible sign of the $b \to c$ decay was seen in early LHCb data~\cite{LHCb-PAPER-2011-040}, in the $\Dbar_1(2420)^0 \to \Dzb\pip\pim$ channel.
In the case that the ground state $D$ meson is reconstructed as a \CP eigenstate, it is possible that decays of $D_1(2420)$ to the $\Dstarp\pim$ and $D(\pip\pim)$ would allow interference between flavour-tagged and untagged $\D$ mesons in the same final state.
However, a full four-body amplitude analysis may be necessary, as there is also a significant contribution from $\Bp \to DK_1^+ \to D\Kp\pip\pim$, which has been used to determine $\gamma$ with a quasi-two-body approach~\cite{LHCb-PAPER-2015-020}.
Further studies will be necessary to establish by how much such an analysis would benefit the sensitivity to $\gamma$.

All $b$~hadron species are produced in $pp$ collisions, and LHCb has recorded large samples of $\Bs$ and $\Lb$ decays.
Hence, possibilities to determine $\gamma$ in $\Bs \to D\Kp\Km$ and $\Lb \to D\proton\Km$ decays can also be considered.
Both of these modes have been observed in LHCb data~\cite{LHCb-PAPER-2012-018,LHCb-PAPER-2013-035,LHCb-PAPER-2013-056}, but with modest yields.
Moreover, a full analysis of $\Bs \to D\Kp\Km$ decays requires tagging of the initial \B meson flavour, which leads to a reduction of sensitivity.
In the case of $\Lb \to D\proton\Km$ decays, the kinematic boundary of the phase space (due to the proton mass) limits overlap between $D\proton$ and $\proton \Km$ amplitudes, and thus it is unclear how much gain in sensitivity may be possible compared to the quasi-two-body analysis.
A detailed study of the amplitude structure of $\Lb \to D\proton\Km$ decays, similar to that recently performed for the related $\Lb \to D\proton\pim$ channel~\cite{LHCb-PAPER-2016-061} will be needed to address this issue.

In summary, $B \to Dhh^\prime$ decays provide many interesting ways to determine $\gamma$, with Dalitz plot analysis methods being particularly sensitive in certain cases.
The results from these methods on the $\Bz \to D\Kstar(892)^0$ mode give competitive sensitivity to those from $\Bp \to D\Kp$, with the precision expected to improve further as results with additional $D$ decay modes become available.
Other $B \to Dhh^\prime$ decays, which have not yet been used to determine $\gamma$, are well worth pursuing, since in addition to helping to improve the overall knowledge of \CP violation, these channels can also provide interesting results in charm meson spectroscopy.

\acknowledgments 
The author thanks the organisers of the CKM2016 conference, and the conveners of WG5, for the invitation to an exciting and well-run meeting.
The work described in these proceedings was supported by the Science and Technology Facilities Council (United Kingdom), and by the European Research Council under FP7.

\addcontentsline{toc}{section}{References}
\setboolean{inbibliography}{true}
\bibliographystyle{LHCb}
\bibliography{references,LHCb-PAPER}

\end{document}